\documentclass[11pt]{article}
\usepackage{graphicx}
\usepackage{amsmath}
\usepackage{epsfig}

\begin{document}

\title{Status of Salerno Laboratory\\
(Measurements in Nuclear Emulsion)\footnote{Presented in The First
International Workshop of Nuclear Emulsion Techniques (12-14 June
1998, Nagoya, Japan), http://flab.phys.nagoya-u.ac.jp/workshop.}}
\author{S. Amendola, E. Barbuto, C. Bozza, C. D'Apolito, \and A. Di Bartolomeo,
M. Funaro, G. Grella, G. Iovane, \and P. Pelosi, G. Romano \\
\\
University of Salerno and INFN, Salerno, Italy} \maketitle

\begin{abstract}
A report on the analysis work in the Salerno Emulsion Laboratory
is presented. It is related to the search for $\nu _{\mu
}\rightarrow \nu _{\tau }$ oscillations in CHORUS experiment, the
calibrations in the WANF (West Area Neutrino Facility) at Cern and
tests and preparation for new experiments.
\end{abstract}

\section{INTRODUCTION}

The analysis work in the Salerno Emulsion Laboratory is running at present
along the following main lines:

\begin{itemize}
\item  the search for $\nu _{\mu }\rightarrow \nu _{\tau }$ oscillations in
CHORUS;

\item  calibrations in the WANF;

\item  tests and preparation for new experiments.
\end{itemize}

The Salerno Group, having developed a new system \cite{rosa} for a
completely automatic system for search and the analysis of
interactions in nuclear emulsions, is also continuing to improve
both hardware and software tools to increase the efficiency and
the speed of the system.

This argument and some of the results obtained in CHORUS have been
presented by E. Barbuto in this workshop.

\section{MEASUREMENT OF MUON FLUX IN THE WANF}

In the West Area Neutrino Facility (WANF) \cite{antonio} at CERN,
a muon-neutrino beam is obtained from the decay of a parallel beam
of pions and kaons of a given sign; the beam is then dumped in
order to remove the charged particles.

The neutrino beam thus obtained is monitored by means of suitable counters
inserted around the primary beam and the target, and inside the dump at
different depths, where the muon flux is measured.

As the neutrino flux is proportional to the number of detected muons, these
measurements allow to center the neutrino beam on its target, and to
estimate its intensity through a Monte Carlo simulation that takes into
account the muon absorption and scattering in the dump.

In addition, this procedure constitutes a real on-line monitoring of the
neutrino beam and allows to detect within a few seconds any fault that could
have occurred along the beam line.

The use of nuclear emulsions is mainly intended to give an
absolute calibration to few of the many Solid State Detectors
(SSDs) inserted in the shielding, that are relatively calibrated
one to the others. This technique was already used in past
\cite{heijne} to calibrate the detectors in the Wide-Band and in
the Narrow-Band-Neutrino beam at CERN.

The present automatic analysis, however, allows to handle much higher
fluxes, with considerably increased statistics and in much shorter times.

\subsection{THE WANF}

The CERN muon-neutrino beam was largely rebuilt in 1992 - 93, and optimized
for a new round of $\nu _{\mu }\rightarrow \nu _{\tau }$ oscillation
experiments (CHORUS and NOMAD).

450 GeV protons from the CERN-SPS are focused onto a primary target in two
spills, 6 msec long, separated in time by 2.7 sec. (fast-slow ejection), one
at the start and one at the end of the flat top. The two spills collect
about 80\% of the primary intensity.

The target is made up of 11 beryllium rods (3 mm in diameter, 10 cm long,
spaced by 9 cm), aligned in a cast aluminum box; the rods are cooled by
means of a closed flux of helium gas.

An aluminum collimator, 2.75 m long, is placed 2.5 m downstream of
the target to remove the wide angle secondaries ($>$ 9 mrad).

After the first two years of successful operation, the station received more
than $2\times 10^{19}$ protons at 450 GeV/c, with a peak intensity of $%
2.8\times 10^{13}$ protons per machine cycle.

Positive (negative) hadrons are produced in the target and are focused
(defocused) by the subsequent beam elements.

The neutrino (antineutrino) beam is generated by positive (negative) hadrons
decays, mainly $\pi ^{+}$, $K^{+}$ $\rightarrow \mu ^{+}$ $+$ $\nu _{\mu }$,
or c.c.

In the case of a neutrino beam, a small contamination (about 5\%) of is
produced by wrong sign hadrons ($\pi ^{-}$, $K^{-}$) present in the selected
beam, while $\nu _{e}$ (about 1\%) $\overline{\nu _{e}}$or (about 0.2\%) are
produced by semileptonic decays of kaons.

A much smaller fraction of $\nu _{\tau }$ (about 10-6) is produced through
the production and the prompt tauonic decay of $D_{s}$.

The beam line of Cern West Area is divided in three principal regions.

\begin{figure}[h!]
\begin{center}
\epsfig{file=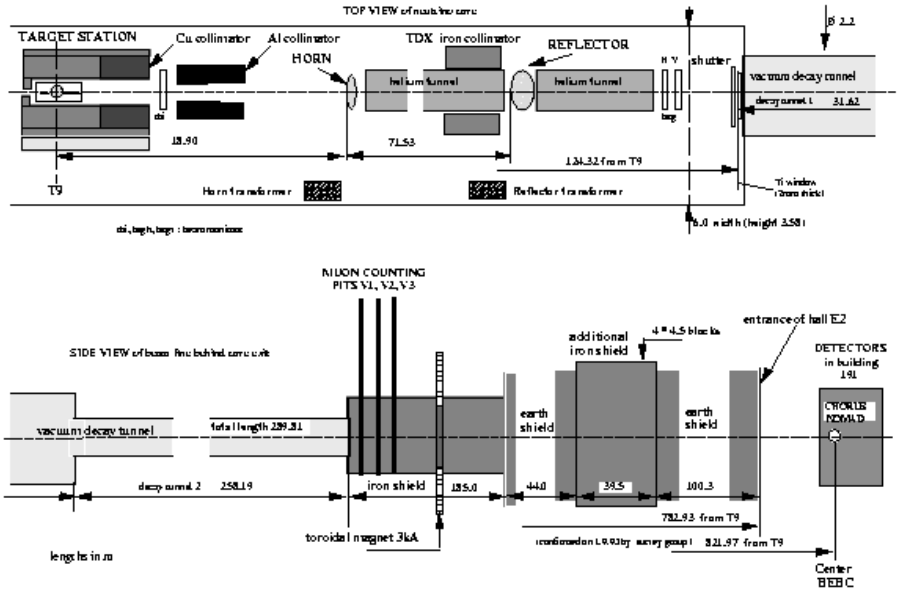,width=12cm,clip=}
\end{center}
\caption{Target and beam line of CERN West Area.} \label{fig1}
\end{figure}

Figure 1 shows schematically, and not to scale, the beam line in
the West Area.

\bigskip

The neutrino cave, 125 m long, contains the target and the
focusing elements. These are the horn and the reflector, a system
of two pulsed, coaxial conductors with cylindrical symmetry,
studied by S. Van der Meer.

The decay tunnel, 290 m long and 60 cm in radius, is evacuated to
a pressure of about 1 torr in order to minimize secondary
interactions. Here, a sizeable fraction of the focused hadrons
decay along a line very close to that of their parents.

The shielding -iron and earth- 400 m long, absorbs the remaining hadrons and
decay muons.

In the shielding there are three pits about 15 m underground, 20, 40 and 60
m downstream of the decay tunnel. In each pit there is an array of 43 Solid
State Detectors (SSD) to monitor the beam.

\subsection{SSDs AND THEIR ABSOLUTE CALIBRATION}

Solide state detectors measure the charge (as an integrated electric
current) produced by crossing particles and need an accurate calibration to
convert their electric signals into charged particles flux measurements.

The absolute calibration is made using the counting provided by
nuclear emulsions. The choice of nuclear emulsions is due to their
very high efficiency and to their excellent two track separation,
that allow the
detection of very intense particle fluxes (up to more than $%
10^{6}~part/cm^{2}$ in our case).

Our work refer to measurements in pit n%
${{}^\circ}$%
3, the most downstream, in which muon flux is less than in the others, and
for this reason, it has been chosen for the emulsion set exposures.

Solid State Detectors (silicon detectors - sensitive surface from 30 up to
200 $mm^{2}$ and thickness from 0.1 up to 1 mm) are placed in boxes on a
rigid support, orthogonal to the beam line.

\begin{figure}[h!]
\begin{center}
\epsfig{file=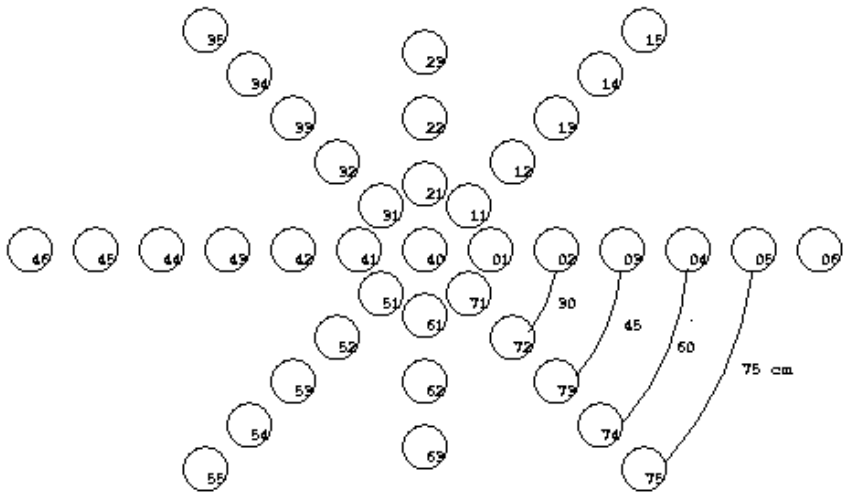,width=12cm,clip=}
\end{center}
\caption{SSDs arrangement in pit n.3.} \label{fig2}
\end{figure}

Figure 2 shows a typical arrangement of the counters.

A particular box with 5 SSDs, the calibration box (Cal.box), that can be
moved in front of the other counters, calibrates them. The calibration
sequence is automatically repeated every 8 hours, by successive movements of
the Cal.box.

Another special box, the reference box (Ref.box), is moved from pit to pit,
and placed in front of the calibration boxes, to calibrate them.

Finally, the absolute calibration of Ref.box, and hence of all the counters,
is performed with emulsion measurements. To be safe, a set of 4 emulsion
plates were used at the same time to calibrate also detector n%
${{}^\circ}$%
40, on the center of beam line, and detector n%
${{}^\circ}$%
2 (during 1994-95) and detector n%
${{}^\circ}$%
42 (during 1996-97).

In the last 4 years, in pit n.3, 17 exposures of emulsion sets were
performed.

Emulsion plates were stuck on the center of a box containing SSDs and
exposed perpendicularly to the beam direction, so that tracks cross the
emulsions nearly perpendicularly and fully automatic scanning can be
successfully performed.

\begin{figure}[h!]
\begin{center}
\epsfig{file=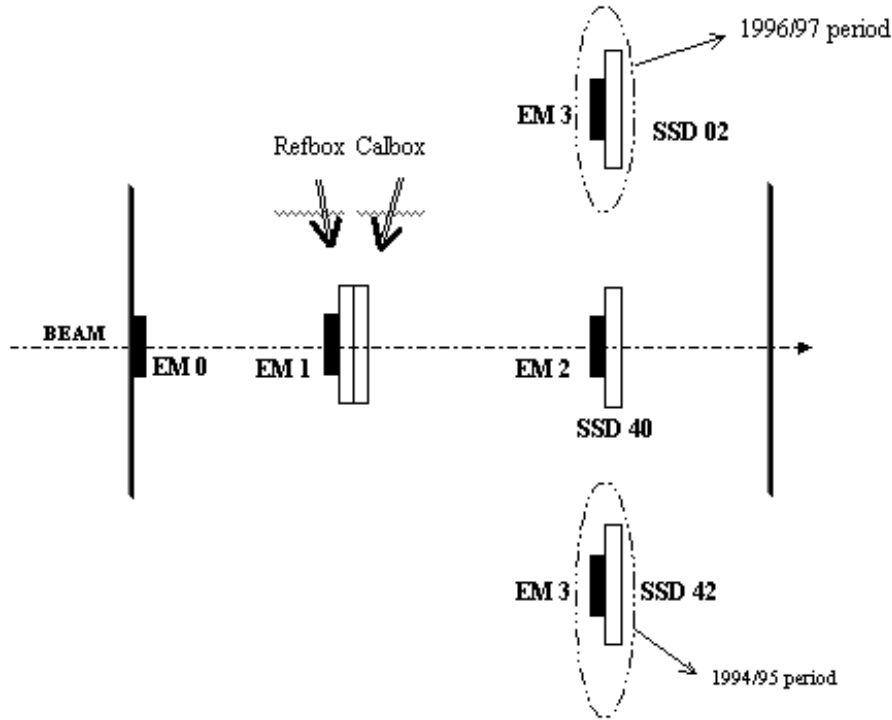,width=12cm,clip=}
\end{center}
\caption{Emulsion plates (black boxes) exposed in pit 3 during
Chorus - Nomad run. Empty boxes represent the Solid State
detectors. The figure is out of scale.} \label{fig3}
\end{figure}

The location of the emulsion plates and their identification
labels are shown in figure 3.

We used 3 or 4 plates ($3\times 4~cm^{2}$) for each set, and two
sets of emulsions were generally exposed the same day, each to one
spill of intensity between $10^{5}~part/cm^{2}$ and $2 \times
10^{6}~part/cm^{2}$, corresponding to $10^{12}\div 10^{13}$
protons on target (p.o.t.).

We have used plates double coated on a 300 $\mu m$ plastic base, with Fuji
or Nikfi gel. The thickness of emulsion pellicles were 100 $\mu m$ (150 $\mu
m$ for the last three exposures).

A total of 66 emulsion plates were exposed; together with each exposed set,
we produced a reference plate that was not exposed in the beam line but was
used for evaluating the background. In order to reduce the background, due
to environment radioactivity, to cosmic rays, etc., the emulsion plates were
poured immediately before and developed soon after the exposure (within a
period, typically, of one week).

We used the automatic scanning, SYSAL (SYstem of SALerno). On each emulsion $%
5\times 5$ fields were scanned (about 0.39 $mm^{2}$), with a gap among
fields to prevent double counting of tracks.

Typically, 35 to 35 tracks/field were found, and a total of about $4\times
10^{5}$ tracks were measured.

A two dimensional image as seen under a $50\times $ magnification is shown
in figure 4, corresponding to a density of about 80 tracks/field.

\begin{figure}[h!]
\begin{center}
\epsfig{file=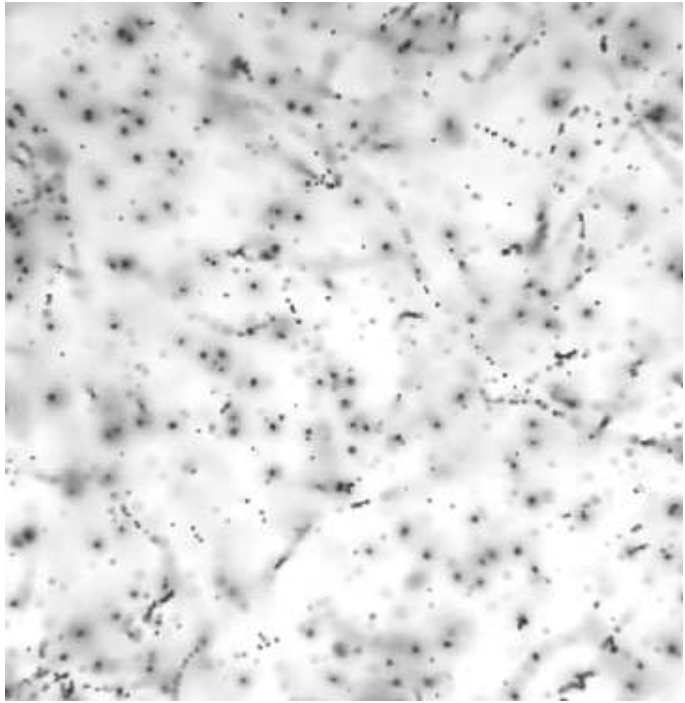,width=9cm,clip=}
\end{center}
\caption{Emulsion layer as seen under a $50\times $ magnification.
Passing - through muons appear as black spots, their tracks being
nearly perpendicular to the plane of the figure.} \label{fig4}
\end{figure}

\subsection{Results}

Tracks that cross the emulsion pellicle from one surface to the other are
called passing - through tracks. In the data taking these tracks are built
up with at least 20 aligned grains, and are the only ones capable to
contribute to the electric signal of SSDs. So, all other tracks present in
emulsion must be excluded from flux evaluation, because they don't reach the
SSDs.

Other tracks reconstructed by SYSAL are:

\begin{itemize}
\item  fake tracks due to association of random grains in emulsions, usually
with few poorly aligned grains;

\item  background tracks, like slow-energy electrons and cosmic rays, mostly
with wide angle;

\item  mixed tracks, poorly aligned grains with an intermediate number of
grains, contribute to both good and background tracks.
\end{itemize}

\begin{figure}[h!]
\begin{center}
\epsfig{file=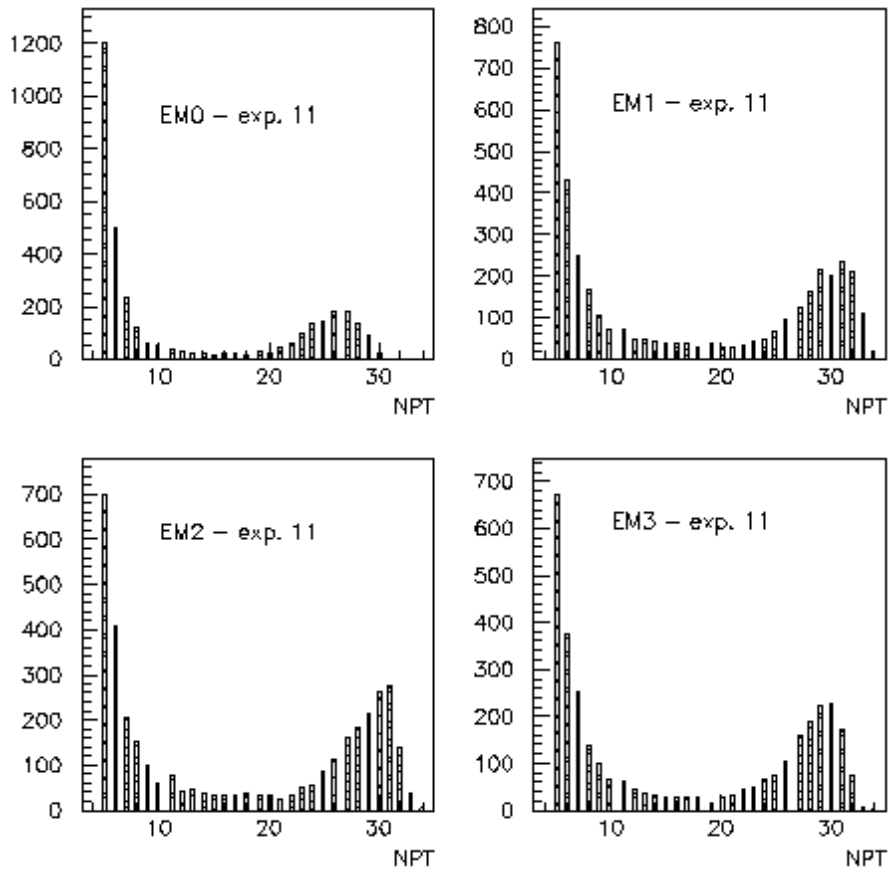,width=12cm,clip=}
\end{center}
\caption{Track grain multiplicity on the top side of the plate of
the $11^{th}$ exposure. The emulsion thickness after development
ranges between about 50 and 60 $\mu m$.} \label{fig5}
\end{figure}

Figure 5 shows the distribution of track grain multiplicity for a
particular exposure.

\bigskip

These plots show a peak at 28 - 30 grains due to passing through tracks; a
flat region includes mixing tracks; another peak with a few grains is due to
fake and background tracks. The cut off at number of points equal to 20, to
calculate muon flux, was chosen in order to reject the latter peak.

\begin{figure}[h!]
\begin{center}
\epsfig{file=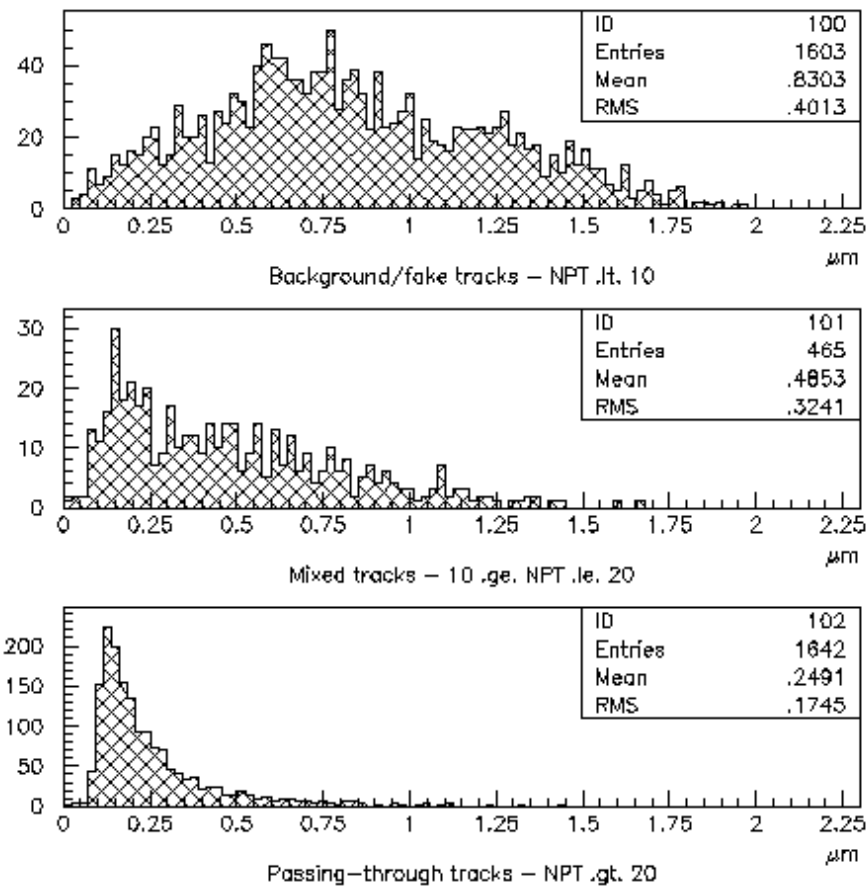,width=12cm,clip=}
\end{center}
\caption{Residuals from a parabolic track fitting.} \label{fig6}
\end{figure}

Figure 6 shows the distributions of the residuals from a parabolic
fitting for the three categories of segments, according to their
grain multiplicity.

\bigskip

To test the reproducibility of our measurements we have rescanned every time
the same plate; such a scanning confirms that the measurements
reproducibility is within a few percent.

For each plate we have also measured polar and azimuthal angle.

\begin{figure}[h!]
\begin{center}
\epsfig{file=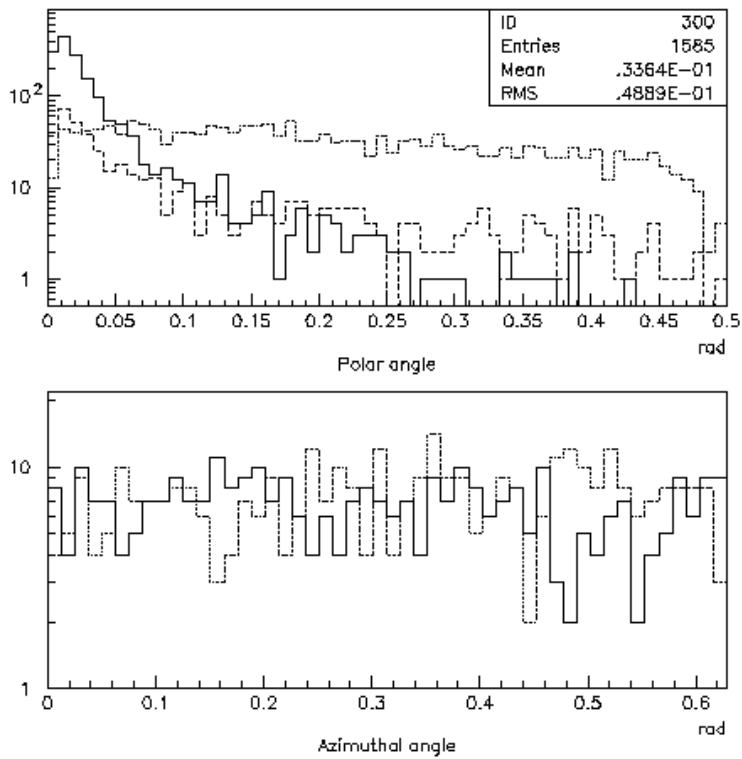,width=10cm,clip=}
\end{center}
\caption{Polar (up) and azimuthal (down) angle measured on an
emulsion pellicle (plate n°1, exposure n° 11) after distortion and
shrinkage corrections for passing - through (continuous line),
mixed (dashed line) and fake/background (dotted line) tracks.}
\label{fig7}
\end{figure}

Figure 7 shows polar and azimuthal angle distributions, after
distortion and shrinkage corrections.

We have studied the correlation between flux from SSDs (electric signal was
converted with the existing calibration) and flux measured on all emulsion
plates. We found a very good linear correlation. A linear correlation is
also confirmed between BCTs\footnote{%
The neutrino flux monitoring system also collects data about the primary
proton beam and the secondary hadron beam from Beam Current Trasformers
(BCTs) upstream of the target.} countings and emulsion measurements.

\begin{figure}[h!]
\begin{center}
\epsfig{file=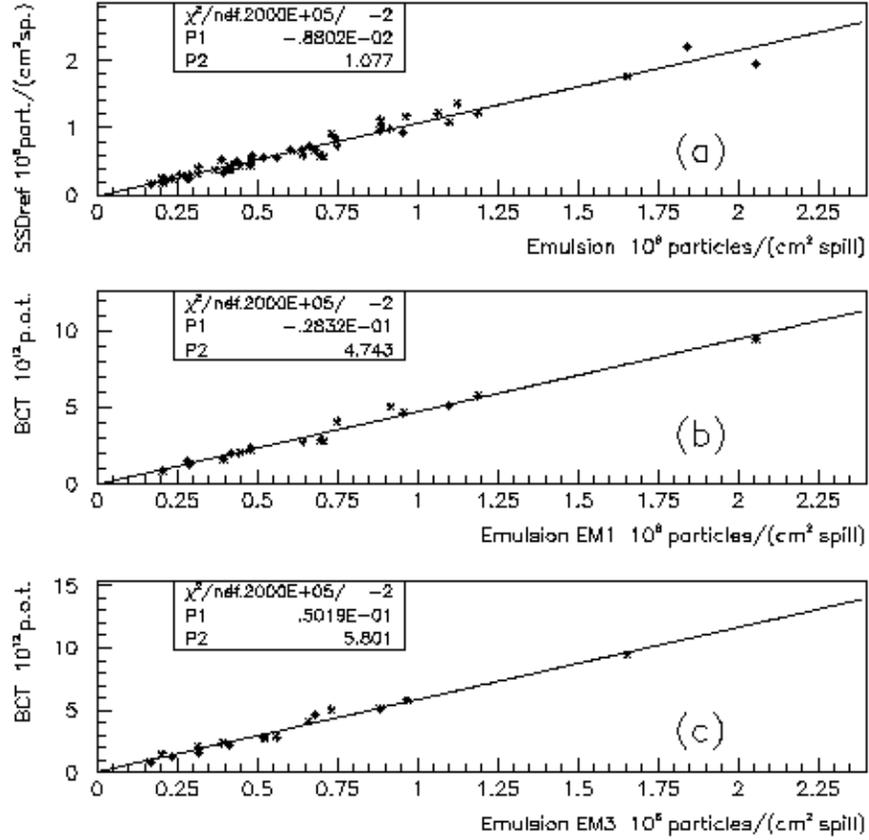,width=12cm,clip=}
\end{center}
\caption{Correlation between the SSD and the BCT counting with
emulsion measurements. In plot (a) the emulsions are related to
the SSDs on which they were exposed; the p.o.t. (proton on target)
are compared with EM1 (plot (b)) and EM2 (plot(c))} \label{fig8}
\end{figure}

These correlations are shown in Figure 8. The correlation between
flux measurements from Ref.box and flux measured from emulsion
plate (Em 1) stuck on Ref.box gives a very good agreement (within
about 1\%).

The highest density of tracks is more than $2\times 10^{6}~tracks/cm^{2}$.

\section{A TEST FOR TOSCA EXPERIMENT}

Sysal was also used to test the set-up of possible, future
experiments. Among these TOSCA \cite{europe}, conceptually similar
to CHORUS, aims to much better detection of charged particles, and
foresees an emulsion target within a magnetic field.

An option for a wide-gap magnet could be the one built for UA1, at present
in NOMAD (maximum field B=0.7 T).

An option for a suitable spectrometer is a Compact Emulsion Tracker (CET), a
set-up already used in the past, for example in the study of high-energy
heavy-ion experiments (EMU 08, 09,16, etc.)

\begin{figure}[h!]
\begin{center}
\epsfig{file=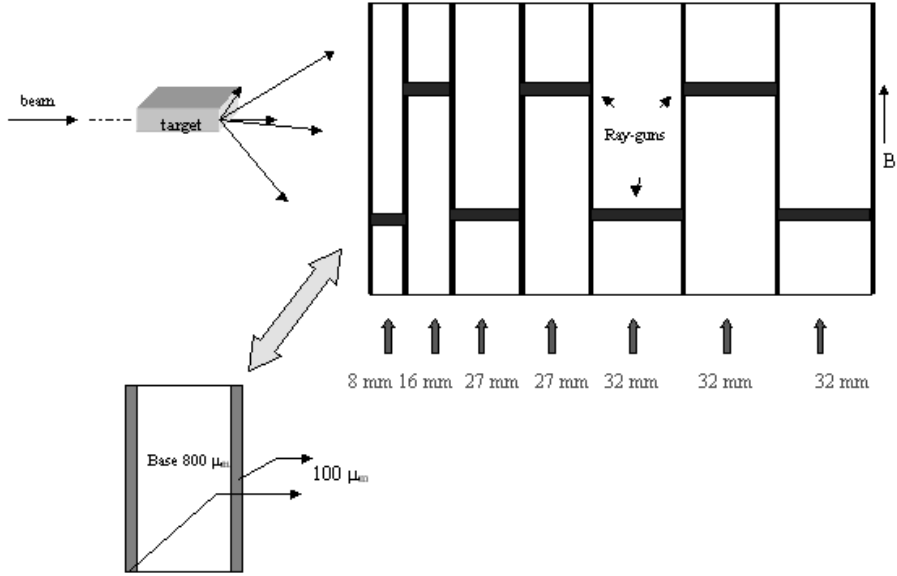,width=12cm,clip=}
\end{center}
\caption{Set up of CETs position. Figure is not in scale}
\label{fig9}
\end{figure}

This set-up (fig. 9) was tested in September 1997 where an
emulsion target 9 cm thick ($3X_{0}$) was followed by silicon
trackers and by a series of 8 thin, parallel emulsion sheets
spaced from 8 to 32 mm over a total length of 180mm.

A 15 GeV beam of pions was used to produce interactions in the target, whose
particles have been traced in the external apparatus.

A first alignment between pairs of CET sheets was realized by means of X-ray
guns (see fig. 9). A better approximation was obtained by measuring
positions and angles of a sample of primary particles, whose individual
identity was recognized from one sheet to the next by comparing the
topological pattern.

\begin{figure}[h!]
\begin{center}
\epsfig{file=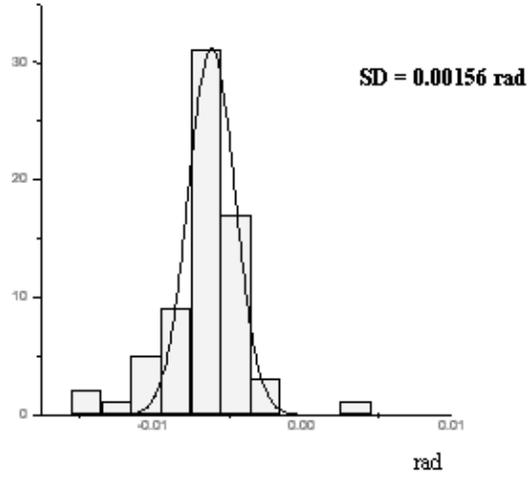,width=7cm,clip=}
\end{center}
\caption{Absolute angular distributions of primaries.}
\label{fig10}
\end{figure}

The angular spread of the beam (fig. 10) was found to be about 1.6 mrad, in
excellent agreement with that expected for a 15 GeV parallel beam after
having crossed a $3X_{0}$ target.

This procedure allowed to place all sheets in a common reference frame by
means of a fit of the rototraslational parameters.

Finally, a random search was performed in the first pellicle for
random tracks out of the beam ($\theta $ $>$ 50 mrad), and thus
some tracks were searched and measured in the following pellicles.

A linear fit was performed for the transverse coordinates as functions of
the longitudinal one: as a result, systematically the residuals on the
bending plane turned out to be about one order of magnitude larger with
respect to those in the non-bending planes, and these are compatible with
the accuracy in locating the pellicles with the location procedure described
above.

A parabolic fit in the bending plane allowed to determine the radius of
curvature, hence the momentum of the particles.

Now, the residuals turn out to be similar to those in the non-bending plane.

Fig. 11 shows a typical example of the results obtained on a
particular track. Table n.1 shows the global results for a set of
tracks.

\begin{equation*}
\begin{tabular}{|c|c|}
\hline $\text{Radius of curvature (m)}$ & $\text{Momentum
(GeV/c)}$ \\ \hline $20.7\pm 3.8$ & $4.3\pm 0.8$ \\ \hline
$-26.3\pm 3.6$ & $-5.5\pm 0.7$ \\ \hline $10.2\pm 0.9$ & $2.1\pm
0.2$ \\ \hline $-23.8\pm 2.4$ & $-5.0\pm 0.5$ \\ \hline $-18.5\pm
2.7$ & $-3.9\pm 0.6$ \\ \hline $17.9\pm 2.0$ & $3.7\pm 0.4$ \\
\hline $-4.17\pm 0.07$ & $-0.87\pm 0.01$ \\ \hline $41.7\pm 6.9$ &
$8.7\pm 1.4$ \\ \hline $-57.5\pm 11.9$ & $-12.1\pm 2.5$ \\ \hline
$-55.6\pm 8.6$ & $-11.7\pm 1.8$ \\ \hline $-2.8\pm 0.09$ &
$-0.58\pm 0.02$ \\ \hline
\end{tabular}
\end{equation*}
\\
Of course, this procedure assumes that the primary particles have
straight trajectories, but their curvature could be easily taken
into account.

In the near future it is planned to repeat these measurements on predicted
tracks, whose momentum has been also measured in the external apparatus.

From these results it is concluded that this set-up allows to
measure momenta with a minimum relative error $\Delta p/p \approx
6\%$, essentially due to multiple scattering in the emulsion
sheets.

The measurement error, that could be possibly lowered, gives at present $%
\Delta p/p \approx 1.4\%\times p$.

\begin{figure}[h]
\begin{center}
\epsfig{file=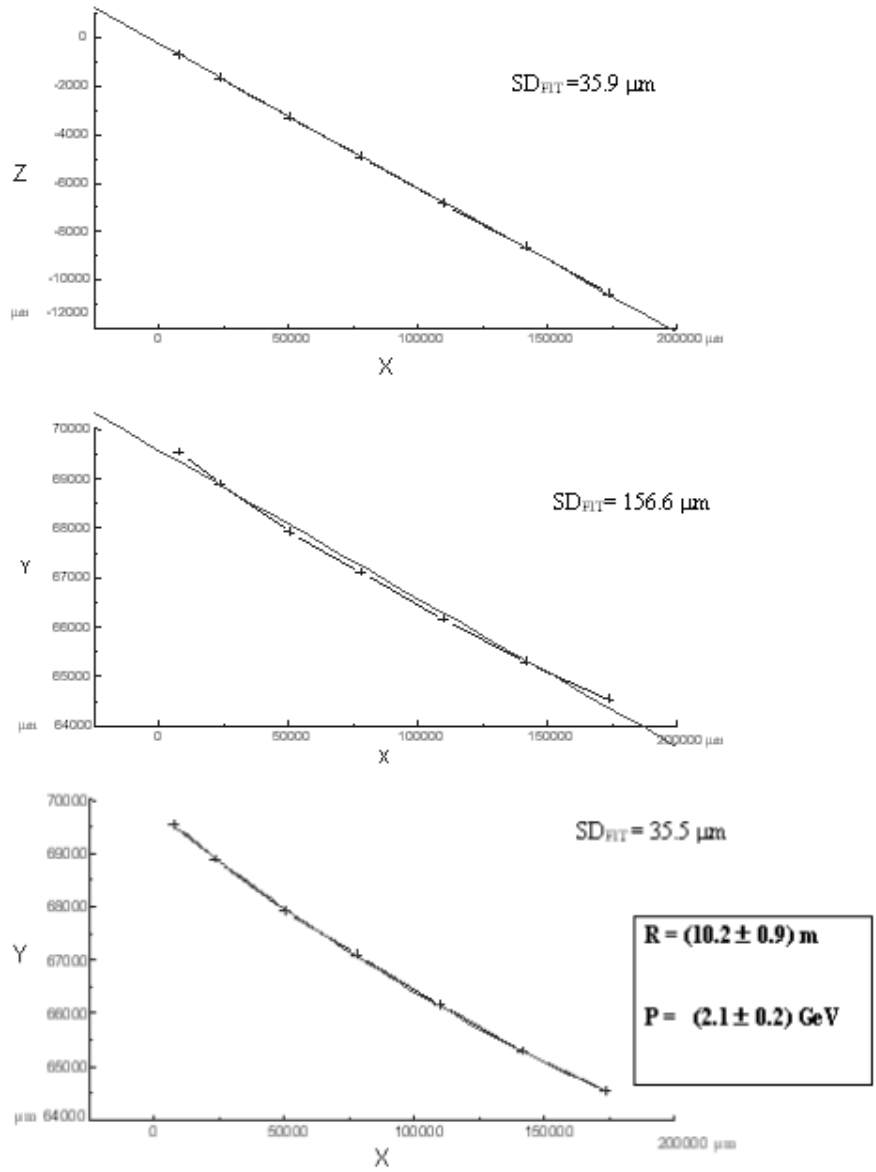,width=12cm,clip=}
\end{center}
\caption{Linear fit of z and y coordinate versus x, parabolic fit
of z coordinate versus x.} \label{fig11}
\end{figure}

\end{document}